# Gravitational field of higher dimensional domain walls in Lyra geometry

F. Rahaman and S. Mandal

Dept. of Mathematics, Jadavpur University, Kolkata-700032, India

E- Mail: farook_rahaman@yahoo.com

## Abstract:

This paper studies Thick domain wall within the framework of Lyra geometry. Their exact solutions are obtained in the background of a five dimensional space-time. The space-time is nonsingular in its both spatial and temporal behavior. The gravitational field of the wall is shown to be attractive in nature.

## Introduction:

Finding a theory that unifies gravity with other forces in nature remains an elusive goal in modern physics. Most efforts in this search have been directed at studying theories in which the number of dimensions of space-time is greater than usual 4-dimensional space-time. For these reasons higher dimensional theory is receiving great attention in both cosmological and non cosmological phenomena [1].

The first study of cosmological implication of higher dimension was made up by Chodos and Detweller [2]. They constructed a Kasner type vacuum space-time in five dimensions, where the extra dimension contracts in course of time although the usual three space expands. Topological objects such as monopoles, strings and domain walls have an important role in the formation of our Universe. Domain walls are formed when the Universe undergoes a series of phase transitions with discrete symmetry being spontaneously broken [3]. Hill, Schramm and Fry [4] have suggested that light domain wall of large thickness may have produced during the late time phase transitions such as those occurring after the decoupling of matter and radiation .

The first solution of Einstein's field equations for the gravitational field produced by a thin planar domain wall was found by Vilenkin in the linear approximation for the gravitational field [5]. Since the linear approximation for the gravitational field breaks down at large distances, this solution could not say anything about the global structure of the gravitational field. After that, there were so many works have been done on domain walls ( general solutions ) both for four and five dimensional space time [6].

While attempting to unify gravitation and electro magnetism in a single space-time geometry, Weyl [7] showed how one can introduce a vector field with an intrinsic geometrical significance. But this theory was not accepted as it was based on non-integrability of length transfer. Lyra [8] proposed a new modifications of Riemannian geometry by introducing a gauge function which removes the non- integrability condition of a vector under parallel transport.

In consecutive investigations Sen [9] and Sen and Dunn [10] proposed a new scalar tensor theory of gravitation and constructed an analog of Einstein field equation based on Lyra's geometry which in which in normal gauge may be written as

$$R_{ab} - \tfrac{1}{2} g_{ab} R + (\tfrac{3}{2}) \varphi_a \varphi_b - \tfrac{3}{4} g_{ab} \varphi_c \varphi^c = - \chi T_{ab} \quad \ldots\ldots(1)$$

where $\phi_a$ is the displacement vector and other symbols have their usual meaning as in Riemannian geometry.

According to Halford [11] the present theory predicts the same effects within observational limits, as far as the classical solar system tests are concerned, as well as tests based on the linearised form of field equations. Soleng [12] has pointed out that the constant displacement field in Lyra's geometry will either include a creation field and be equal to Hoyle's creation field cosmology or contain a special vacuum field which together with the gauge vector term may be considered as a cosmological term

Subsequent investigations were done by several authors in scalar tensor theory and cosmology within the framework of Lyra geometry [13].

In recent, Farook has studied some topological defects namely monopole, cosmic string and domain wall in the framework of Lyra geometry [14].

In this work we shall deal with thick domain wall in higher dimensions with both time and space dependent displacement vectors based on Lyra geometry in normal gauge i.e. displacement vector

$$\phi_i = ( \beta (z,t) ,0,0,0,0) \quad \ldots\ldots\ldots(2)$$

and look forward whether the domain wall shows any significant properties due to introduction of the gauge field in the Riemannian geometry .

Thick domain walls are characterized by the energy momentum tensors :

$$T_{ij} = \rho ( g_{ij} + w_i w_j ) + p w_i w_j \; ; \; w_i w^i = -1$$

where $\rho$ is the energy density of the wall, p is the pressure in the direction normal to the plane of the wall and $w_i$ is a unit space like vector in the same direction .

We will discuss two models: The stress tensor $T_\psi^\psi$ corresponding to fifth dimension is assumed to be zero in the first case and equal to $T_x^x = T_y^y$ in the second case.

## 2. An over view on Lyra geometry:

Lyra geometry is a modification of Riemannian geometry, which bears a close similarity to Weyl's geometry.

Lyra defined the displacement vector PF between two neighboring points $P(x^i)$ and $F(x^i + d\, x^i)$ by its components $A\, dx^i$ is a gauge function A together form a reference system ($A, x^i$).

The transformation to a new reference system ($x^i, x_1^i$) are given by

$$A_1 = A_1(A, x^i), \quad x_1^i = x_1^i(x^i)$$

with $(\partial A_1 / \partial A) \neq 0$ and the Jacobian $|\partial x_1^i / \partial x^i| \neq 0$

The connections is taken as $\quad {^*\Gamma_{bc}^a} = A^{-1} \Gamma_{bc}^a - \tfrac{1}{2}(\delta_b^a \varphi_c + \delta_c^a \varphi_b - g_{cb}\varphi^a)$ ....(3)

where the $\Gamma_{bc}^a$ are defined in terms of the metric tensor $g_{ab}$ as in Riemannian geometry and $\varphi_a$ is a displacement vector field. Lyra [8] and Sen [9,10] have shown that in any general system the vector field quantities $\varphi_a$ arise as a natural consequence of the introduction of the gauge function A into the structure less manifold. ${^*\Gamma_{bc}^a}$ are symmetric in their lower two indices.

The metric in Lyra's geometry is given by

$$ds^2 = A^2\, g_{ij}\, x^i\, x^j \qquad\qquad \ldots\ldots(4)$$

and is invariant both co ordinate and gauge transformations.
The infinitesimal parallel transport of a vector $\xi^a$ is given by

$$d\, \xi^a = -\hat{\Gamma}_{bc}^a\, \xi^b\, A\, d\, x^c \qquad\qquad \ldots\ldots(5)$$

where $\quad \hat{\Gamma}_{bc}^a = {^*\Gamma_{bc}^a} - \tfrac{1}{2}\delta_b^a\,\varphi_c \qquad\qquad \ldots\ldots(6)$

The $\hat{\Gamma}_{bc}^a$ are not symmetyric in b and c.
The length of a vector does not change under parallel transport.
The curvature tensor is defined by

$${^*R^a}_{bcd} = A^{-2}[-(A\hat{\Gamma}_{bc}^a)_{,d} + (A\hat{\Gamma}_{bd}^a)_{,c} - A^2(\hat{\Gamma}_{bc}^e\,\hat{\Gamma}_{ed}^a - \hat{\Gamma}_{bc}^e\,\hat{\Gamma}_{ce}^a)] \qquad \ldots(7)$$

The curvature scalar, obtained by contraction of eq.(7) is

$${^*R} = A^{-2} R + 3 A^{-1}\varphi^a_{;a} + (3/2)\,\varphi^a \varphi_a + 2 A^{-1}(\log A^2)_{,a}\,\varphi^a \qquad \ldots\ldots(8)$$

The invariant volume integral is given by

$$I = \int L\, \sqrt{(-g)}\, A^4 d^4x \qquad\qquad \ldots\ldots(9)$$

where $d^4x$ is the volume element and L is a scalar invariant.

If we use a normal gauge i.e. A = 1 and L = *R in eq.(9), then eqs.(8) and (9) becomes respectively

$$*R = R + 3\varphi^a_{;a} + (3/2) \varphi^a \varphi_a \qquad \ldots\ldots(10)$$

$$I = \int *R \sqrt{(-g)} d^4x \qquad \ldots\ldots(11)$$

The field equations are obtained from the variational principle

$$\delta(I + J) = 0 \qquad \ldots\ldots(12)$$

where I is given as by eq.(11) and J is related to the Lagrangian density **L** of matter by

$$J = \int L \sqrt{(-g)} d^4x \qquad \ldots\ldots(13)$$

The field equations are thus ( using $\chi = 8\pi G / c^4$ )

$$R_{ik} - \tfrac{1}{2} g_{ik} R + (3/2) \phi_i \phi_k - \tfrac{3}{4} g_{ik} \phi_m \phi^m = -\chi T_{ik} \qquad \ldots(14)$$

## 3. The models and the Basic equations:

The general five dimensional plane symmetric metric can be expressed in the form.

$$ds^2 = e^A (dt^2 - dz^2) - e^C (dx^2 + dy^2) - e^\mu d\psi^2 \qquad \ldots\ldots(15)$$

where $A = A(z,t)$; $C = C(z,t)$; $\mu = \mu(z,t)$.

The energy stress components in comoving coordinates for the domain wall under consideration here are given by

$$T_t^t = T_x^x = T_y^y = \rho \; ; \; T_z^z = -p, \; T_\psi^\psi = p_1 \text{ and } T_t^z = 0 \qquad \ldots\ldots(16)$$

$\rho$ is the energy density of the wall, which is equal to the tension along x and y directions in the plane of the wall, p is the pressure along z-direction. The stress tensor $T_\psi^\psi = p_1$ corresponding to fifth dimension is assumed to be zero in the first case and equal to $T_x^x = T_y^y$ in the second case.

The field equation (1) for the metric (15) reduces to
$$\tfrac{1}{4} e^{-A} [-4 C^{11} - 2\mu^{11} - 3 (C^1)^2 - (\mu^1 - A^1)(2C^1 + \mu^1)$$
$$+ C^{*2} + 2 C^* (A^* + \mu^*)] - \tfrac{3}{4} \beta^2 e^{-A} = \chi \rho \qquad \ldots(17)$$

$$\tfrac{1}{4} e^{-A} [4 C^{**} + 2\mu^{**} + 3 C^{*2} + (\mu^* - A^*)(2C^* + \mu^*)$$

$$- (C^1)^2 - A^1 \mu^1 - 2 C^1(A^1 - \mu^1)] + \tfrac{3}{4} \beta^2 e^{-A} = - \chi p \quad \ldots\ldots(18)$$

$$\tfrac{1}{4} e^{-A} [-2(A^{11} + C^{11} + \mu^{11}) - (C^1)^2 - (\mu^1)^2 - C^1 \mu^1$$
$$+ 2(C^{**} + A^{**} + \mu^{**}) + C^{*2} + C^* \mu^* + \mu^{*2}] + \tfrac{3}{4} \beta^2 e^{-A} = \chi \rho \quad \ldots(19)$$

$$\tfrac{1}{4} e^{-A} [-4 C^{11} - 2 A^{11} - 3 (C^1)^2 + 4 C^{**} + 2 A^{**} + 3 C^{*2}] + \tfrac{3}{4} \beta^2 e^{-A} = 0 \quad \ldots(20)$$

$$\tfrac{1}{2} [-C^{*1} + C^*(A^1 - C^1) + C^1 A^* - \mu^{*1} + A^1 \mu^* + \mu^1 A^* - \mu^1 \mu^*] = 0 \quad \ldots(21)$$

Further the conservation equations $[G_{ik} + (3/2) \phi_i \phi_k - \tfrac{3}{4} g_{ik} \phi_m \phi^m + \chi T_{ik}]_{;k} = 0$ reduce

$$\chi \rho^* + (3/2) \beta^* \beta + [\chi(\rho+p) + \tfrac{3}{4} \beta^2][A^* + 2C^*] + \tfrac{1}{2} \mu^* [\chi(\rho+p_1) + \tfrac{3}{4}\beta^2] = 0 \quad \ldots(22)$$

$$\chi p^1 + (3/2) \beta^1 \beta + [\chi(\rho+p) + \tfrac{3}{4} \beta^2][A^1 + 2C^1] + \tfrac{1}{2} \mu^1 [\chi(\rho+p_1) + \tfrac{3}{4}\beta^2] = 0 \quad \ldots(23)$$

[ '*' and '1' are differentiations w.r.t. t and z respectively ]

# 4. Solutions:

## Case-I:

To solve the field equations, we shall assume the separable form of the metric coefficients as follows:

$$A = A_1(z) + A_2(t) \,;\, C = C_1(z) + C_2(t) \,;\, \mu = \mu_1(z) + \mu_2(t) \,; \quad \ldots(24)$$

From eq.(21), by using the separable form, we get

$$C_2^*(A_1^1 - C_1^1) + \mu_2^*(A_1^1 - \mu_1^1) + A_2^*(C_1^1 + \mu_1^1) = 0 \quad \ldots(25)$$

This leads to a possible choice

$$A_1^1 = C_1^1 \quad \ldots(26)$$

so that
$$(\mu_2^* / A_2^*) = (C_1^1 + \mu_1^1) / (-A_1^1 + \mu_1^1) = m \text{ ( separation constant )} \ldots(27)$$

Thus we get,
$$\mu_2 = m A_2 \quad \ldots(28)$$
and
$$A_1 = d \mu_1 \quad \ldots(29)$$

where $d = (m-1)/(m+1)$.

Now we introduce here an assumption

$$C_2(t) = a\, A_2(t) \quad \text{( where a is an arbitrary constant )} \quad \ldots(30)$$

Taking now the following combination of equations (17) – (19) + 2.(20), we get

$$6d\, \mu_1^{11} + 3d^2 (\mu_1^1)^2 = (2a - 2m) A_2^{**} + (3a^2 + am + 2a - m^2) C_2^{*2} = n \quad \ldots(31)$$

( n being the separation constant )

Solving eq.(31), we get

$$\mu_1 = \ln [\cosh(\tfrac{1}{2} dg_1 z)]^{2/d}, \text{ where } g_1^2 = [n/3d^2] \quad \ldots(32)$$

For time part, we get

$$A_2 = \ln [\cosh(\tfrac{1}{2} vg_2 t)]^{1/v} \quad \ldots(33)$$

where $g_2^2 = [n/(3a^2 + am + 2a - m^2)]$ and $v = [(3a^2 + am + 2a - m^2)/2(a-m)]$

So finally the complete solutions for the metric coefficients may be expressed in the form

$$\left.\begin{array}{l} e^A = [\cosh(vg_2 t)]^{1/v} \cdot [\cosh(\tfrac{1}{2} dg_1 z)]^2, \\[4pt] e^C = [\cosh(vg_2 t)]^{a/v} \cdot [\cosh(\tfrac{1}{2} dg_1 z)]^2, \\[4pt] e^\mu = [\cosh(vg_2 t)]^{m/v} \cdot [\cosh(\tfrac{1}{2} dg_1 z)]^{2/d}, \end{array}\right\} \quad \ldots(34)$$

The energy density $\rho$ and the pressure p in the z-direction within the wall are

$$\chi \rho = \tfrac{1}{4} e^{-A} [(16a + 8) g_2^2 v \{\operatorname{sech}(vg_2 t)\}^2 + (13a^2 + 2a + 2am) g_2^2 \{\tanh(vg_2 t)\}^2$$
$$- (14d + 1) g_1^2 d\{\operatorname{sech}(\tfrac{1}{2} dg_1 z)\}^2 + (17d^2 - d - 1) g_1^2 \{\tanh(\tfrac{1}{2} dg_1 z)\}^2] \quad \ldots(35)$$

$$-\chi p = \tfrac{1}{4} e^{-A} [(2m - 2) g_2^2 v\{\operatorname{sech}(vg_2 t)\}^2 + \{(m-1)(2a + d) - a^2\} g_2^2 \{\tanh(vg_2 t)\}^2$$
$$+ (12d^2 g_1^2)\{\operatorname{sech}(\tfrac{1}{2} dg_1 z)\}^2 + (9d^2 + d) g_1^2 \{\tanh(\tfrac{1}{2} dg_1 z)\}^2] \quad \ldots(36)$$

Here $\beta^2 (z,t)$ takes the following form

$$\tfrac{3}{4} \beta^2 = 3 g_1^2 d^2 \{\operatorname{sech}(\tfrac{1}{2} dg_1 z)\}^2 + 3 d^2 g_1^2 \{\tanh(\tfrac{1}{2} dg_1 z)\}^2$$
$$- (4a + 2) g_2^2 v\{\operatorname{sech}(vg_2 t)\}^2 - 3a^2 g_2^2 \{\tanh(vg_2 t)\}^2 \quad \ldots(37)$$

## Case II:

Here we construct another model of a domain wall in five dimensional space time.

We assume here $T_t^t = T_x^x = T_y^y = T_\psi^\psi = \rho$.

In view of the above forms of energy stress tensors and using the field equations, we find the following solutions:

$$\left. \begin{array}{l} e^A = [\cosh(fg_4t)]^{1/f} \cdot [\cosh(ug_3z)]^{d/u}, \\[6pt] e^C = [\cosh(fg_4t)]^{b/f} \cdot [\cosh(ug_3z)]^{d/u}, \\[6pt] e^\mu = [\cosh(fg_4t)]^{m/f} \cdot [\cosh(ug_3z)]^{1/u}, \end{array} \right\} \quad \ldots(38)$$

where $u = [(2d^2 - d - 1)/(2d - 2)]$ ; $g_3^2 = [q/(2d^2 - d - 1)]$ ;

$f = [(2b^2 - m^2 - mb)/(2b - 2m)]$ ; $g_4^2 = [q/(2b^2 - m^2 - mb)]$

[ here we also use an assumption $C_2^* = b\,A_2^*$ , b is arbitrary constant and q is separation constant ]

Here we get the following expressions of $\rho$, p and $\beta^2$ as

$$\chi\rho = \tfrac{1}{4} e^{-A}[(2m+2b+2)g_4^2 f\{\operatorname{sech}(fg_4t)\}^2 + \{2b^2+m^2+mb+2b(1+m)\}g_4^2\{\tanh(fg_4t)\}^2$$
$$+ (3d^2 g_3^2)\{\operatorname{sech}(ug_3z)\}^2 - (8d+4)ug_3^2\{\tanh(ug_3z)\}^2] \quad \ldots(39)$$

$$-\chi p = \tfrac{1}{4}e^{-A}[(2m+4)g_4^2 f\{\operatorname{sech}(fg_4t)\}^2 + \{(m-1)(2b+d)+b^2+4b(1+m) -$$
$$2m(m+b)\}g_4^2\{\tanh(fg_4t)\}^2 + (3d^2+5D)g_3^2\{\tanh(ug_3z)\}^2] \quad \ldots(40)$$

$$\tfrac{3}{4}\beta^2 = -3d^2 g_3^2\{\tanh(ug_3z)\}^2 - (2b+2m+2)g_4^2 f\{\operatorname{sech}(fg_4t)\}^2 +$$
$$(2b+mb-m^2)g_4^2\{\tanh(fg_4t)\}^2 \quad \ldots(41)$$

## 5. Discussions:

From the results given above, it is evident that at any instant the domain wall density ρ as well as pressure p in the perpendicular direction decrease on sides of the away from the symmetry plane and both vanish as $z \to \pm \infty$.

The general expression for the three space volume is given by

$$|g_3|^{1/2} = [\cosh(\tfrac{1}{2} dg_1 z)]^6 \cdot [\cosh(vg_2 t)]^{(1+2a)/v} \qquad \ldots(42)$$

Thus the temporal behavior would be $|g_3|^{1/2} \sim [\cosh(vg_2 t)]^{(1+2a)/v}$. ...(43)

If $v < 0$, then the three space collapses with the extra dimension. In this process one meets a sheet like singularity with expansions along the plane of the wall but collapses along z-direction. On the other hand when $v > 0$, there are expansions along z-direction with collapses in the plane.

Similar results exist for model two.

One interesting feature in our higher dimensional thick domain wall in Lyra geometry that the ratio of ρ and p is not constant. But in higher dimension domain wall in general relativity, one can see that the ratio of ρ and p is constant [ cf. Banerjee et al (1998) [6]]

Thus we see an important difference between domain wall in Lyra geometry with the classical result.

Another aspect of the domain wall is the effect on a test particle in its gravitational field.

The repulsive and attractive character of the thick domain wall can be discussed by either studying the time like geodesics in the space-time or analyzing the acceleration of an observer who is rest relative to the wall. In case-I, let us consider an observer with four velocity given by

$$V_i = \cosh(\tfrac{1}{2} dg_1 z)\cosh^{1/2v}(vg_2 t)\, \delta_i^t$$

Then we obtain the acceleration vector $A^i$ as

$$A^i = V^i{}_{;k} V^k = (\tfrac{1}{2} dg_1)\tanh(\tfrac{1}{2} dg_1 z)[\mathrm{sech}(vg_2 t)]^{1/v} \cdot [\mathrm{sech}(\tfrac{1}{2} dg_1 z)]^2 \delta_z^i. \quad \ldots(45)$$

It is evident that $A^z$ is positive.

It follows that an observer who wishes to remain stationary with respect to the wall must accelerate away from the wall. In other words the wall exhibits a attractive nature to the observer. Similar conclusion can be drawn for the domain wall solutions in case-II.

Hence our result is similar to Banerjee et al [6] domain wall.

Thus our domain walls in Lyra geometry exhibit peculiar (some properties similar and some properties contrary to Banerjee et al domain wall) features. Before ending up our discussions, we are surprising to note that the displacement vector still exists after infinite times. So for future work it will be interesting to study different topological defects within the framework of Lyra geometry.

# Acknowledgement:

We are to grateful Prof. S Chakraborty and Prof. A.Banerjee for helpful discussion. F.R. is thankful to IUCAA for providing the research facilities.